\documentclass[twocolumn,showpacs,preprintnumbers,amsmath,amssymb]{revtex4}

\usepackage{graphicx}
\usepackage{dcolumn}
\usepackage{bm}

\begin{document}


\title{Fiber lasers generating radially and azimuthally polarized light}

\author{Moti Fridman, Nir Davidson and Asher A. Friesem}

\address{Dept. of Physics of Complex System, Weizmann Institute of Science, Rehovo 76100, Israel}

\author{Galina Machavariani}

\address{Soreq Nuclear Research Center, Electro-Optics Division, Yavne 81800, Israel}


\begin{abstract}
A simple, robust, and efficient method to produce either radially
or azimuthally polarized output beam from a fiber laser is
presented. Experimental results reveal that polarization purity of
90\% or better can be obtained.
\end{abstract}

\maketitle

\newpage

Radially and azimuthally polarized laser beams have unique
properties and symmetries. For example, radially polarized light
under tight focusing conditions, produces strong longitudinal
electric field at the focus, while the azimuthally polarized light
produces strong longitudinal magnetic field~\cite{Focus1}. Such
properties are advantageous for various applications including
microscopy~\cite{Microscopy}, material
processing~\cite{LaserProssessing, Drill}, trapping and
acceleration of particles~\cite{Trap1, Trap2} and laser light
amplifications~\cite{GalinaAmp}. Moreover, since the cross-section
intensity distribution of radially and azimuthally polarized light
have a doughnut shape, their peak intensity is significantly lower
than that of a Gaussian light, so that non-linear and damage
effects are reduced. Also, radially and azimuthally polarized
light can propagate inside a non polarization maintaining
multi-mode fiber with much less effects from birefringence than
linearly polarized light~\cite{BandGap, Grosjean1, Volpe,
Grosjean2}.

Over the years, many methods have been developed in order to
generate radially and azimuthally polarized light with lasers.
With solid state lasers they involved phase elements~\cite{Oron},
computer generated holograms~\cite{ComputerHologram} and spatially
variable retarder~\cite{SVR2008}. While in fiber lasers they
involved an intra-cavity axicon~\cite{ShirakwaRadialFiber1} and an
intra-cavity dual conical element~\cite{ShirakwaRadialFiber2} to
produce only radially polarized light; unfortunately, these
methods are highly sensitive and have low polarization purity.

In this letter, we present a highly efficient method for obtaining
either radial or azimuthal polarization directly from a fiber
laser. This method involves the introduction of strong losses to
an unwanted polarization inside the fiber laser cavity by means of
a spatially variable retarder~\cite{GalinaSVR} and a thin film
polarizer. While this method has been successfully demonstrated
with solid-state lasers~\cite{SVR2008}, it was not clear whether
it would be effective with fiber lasers that have relatively
strong nonlinearities, high gain, and birefringence. Especially,
with fiber lasers which include multimode fibers and are
non-polarization maintaining fibers.

\begin{figure}[h]
\centerline{\includegraphics[width=8cm]{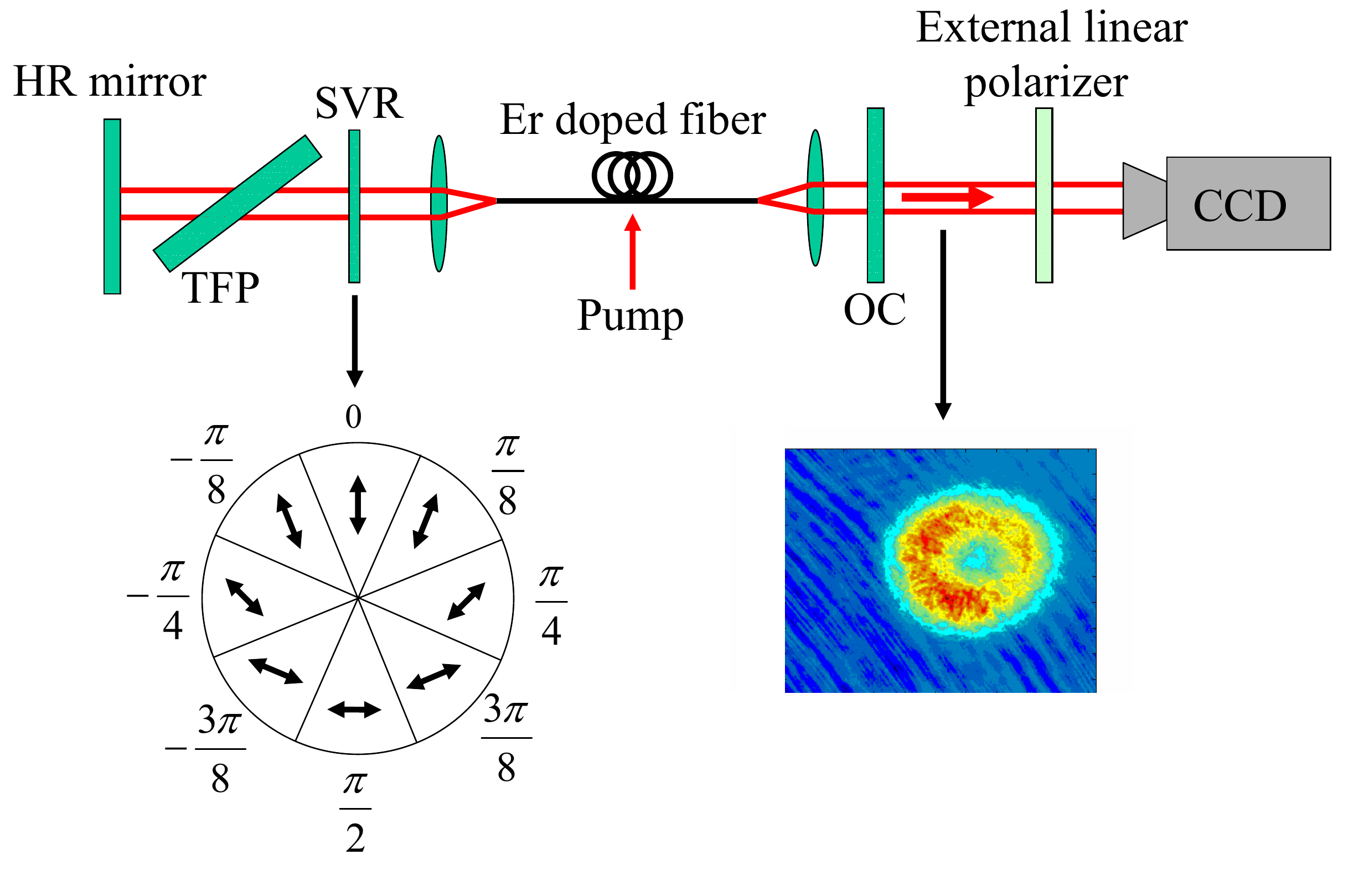}}\caption{Basic
configuration for a fiber laser generating radially and
azimuthally polarized light. TFP - thin film polarizer, OC -
output coupler, HR - high reflection mirror, SVR - spatially
variable retarder. Lower right inset shows the typical
experimental intensity distribution of the output beam with radial
polarization. Lower left inset shows the orientation of the 8
sectors of the SVR.}  \label {setup}
\end{figure}

The basic configuration for generating radially and azimuthally
polarized light in a fiber laser, is schematically presented in
Fig. 1. It includes ten meter long non-polarization maintaining
Erbium-doped fiber. The fiber can support few modes and has a
numerical aperture of 0.07, core diameter of $22 \mu m$ and
cladding diameter of $125 \mu m$. The fiber is side pumped from
both directions with two $911nm$ multi-mode diode lasers of
$100mW$ each. The two fiber ends are cleaved at $8^{\circ}$ to
suppress any reflection back into the fiber. The light emerging
from one end of the fiber is collimated and propagates toward a
4\% reflecting output coupler (OC), while the light emerging from
the other end is collimated and propagates toward the highly
reflecting (HR) back mirror. A spatially variable retarder
(SVR)~\cite{GalinaSVR} and a thin film polarizer (TFP) are
inserted between the collimating lens and the back mirror. The TFP
reflects the S polarized light out of the laser so only the P
polarized light is reflected back towards the SVR. The SVR is
comprised of eight sectors each having $\lambda / 2$ retardation
in different orientations, where the direction of the slow axis of
each retardation plate is denoted by an arrow in Fig. 1. When the
SVR is oriented at $0^{\circ}$, it converts a linearly polarized
light to radially polarized light that is reflected back into the
fiber, so only radially polarized light can exist inside the fiber
laser cavity. When the SVR is oriented at $45^{\circ}$ it converts
the linearly polarized light to azimuthally polarized light so
only azimuthally polarized light can exist in the fiber laser
cavity. The SVR is suitable for high power and has been
successfully tested up to the kW level~\cite{GalinaSVRHigh}. Since
both radially and azimuthally polarized lights have an inherent
cylindrical symmetry similar to the fiber, they suffer much less
from birefringence~\cite{BandGap, Grosjean1, Volpe, Grosjean2}.

\begin{figure}[h]
\centerline{\includegraphics[width=8cm]{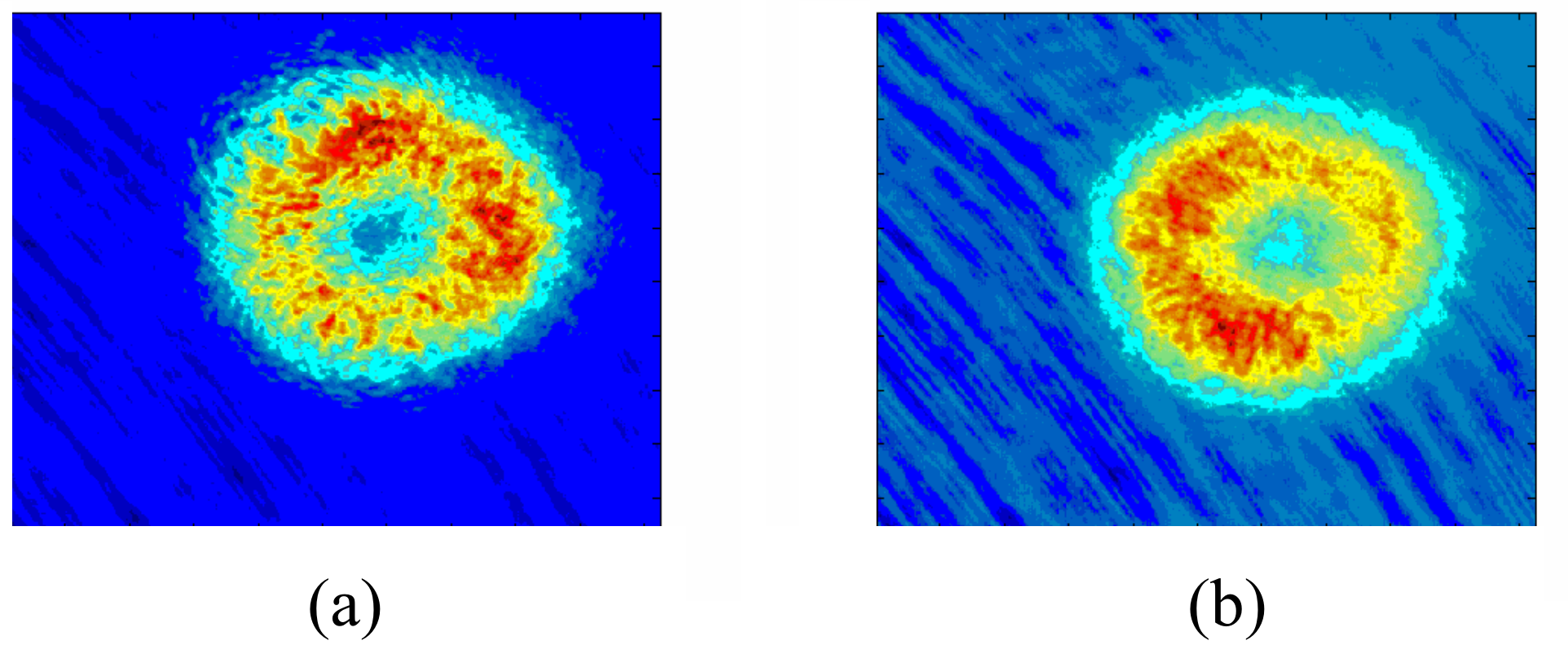}}\caption{Typical
cross-sections of the intensity distributions of the output beam
from the fiber laser. (a) Radially polarized light; (b)
azimuthally polarized light. } \label {doughnut}
\end{figure}

In our experiments we first set the SVR at $0^{\circ}$ orientation
so as to obtain radial polarization, and then rotated the SVR by
$45^{\circ}$ to obtain azimuthal polarization. In both cases the
output power of the laser was about $45mW$. The light emerging
from the fiber laser was then imaged onto a CCD camera. The
detected cross-section intensity distribution of the light for
both cases are presented in Fig. 2. As evident, the expected
doughnut shaped distributions were obtained, and the beam quality
of the outgoing beam was measured to be $M^2=2.05$. In order to
determine the exact polarization distribution of the output light
for both cases, we inserted a linear polarizer in front of the CCD
camera, rotated it to several discrete orientations and detected
the resulting intensity distributions.

\begin{figure}[h]
\centerline{\includegraphics[width=8cm]{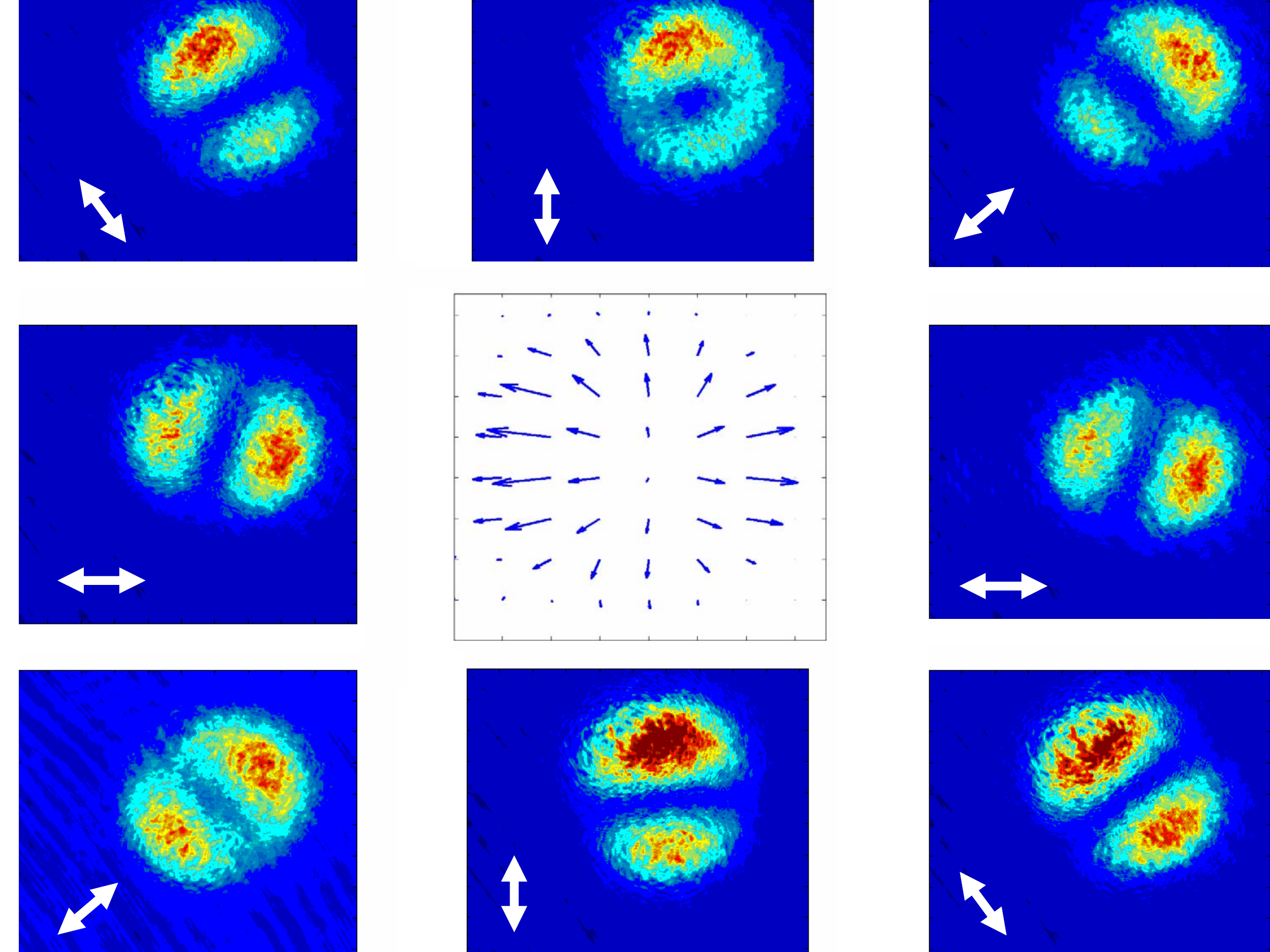}}\caption{Representative
cross-sections of the intensity distributions of the output beam
after the external linear polarizer in eight different
orientations when the SVR is oriented at $0^{\circ}$. At the
center, the resulting main axis of the local polarization
ellipse.} \label {radial}
\end{figure}

The results for eight representative orientations are presented in
Figs. 3 and 4. Figure 3 shows the results when the SVR is oriented
at $0^{\circ}$. As evident from the detected distributions, the
orientation of the common bisector of the two lobes is parallel to
the direction of the external linear polarizer (denoted by the
arrows), indicating radial polarization. Using these detected
distributions, we calculated the local polarization at any point
across the output beam~\cite{Oron}. The results are shown in the
center of Fig. 3, where the arrows indicate the direction of the
main axis of the local polarization ellipse. Using these results
we determined that the radial polarization purity, defined as the
amount of light that would pass through a radial polarizer, is
92\%.

\begin{figure}[h]
\centerline{\includegraphics[width=8cm]{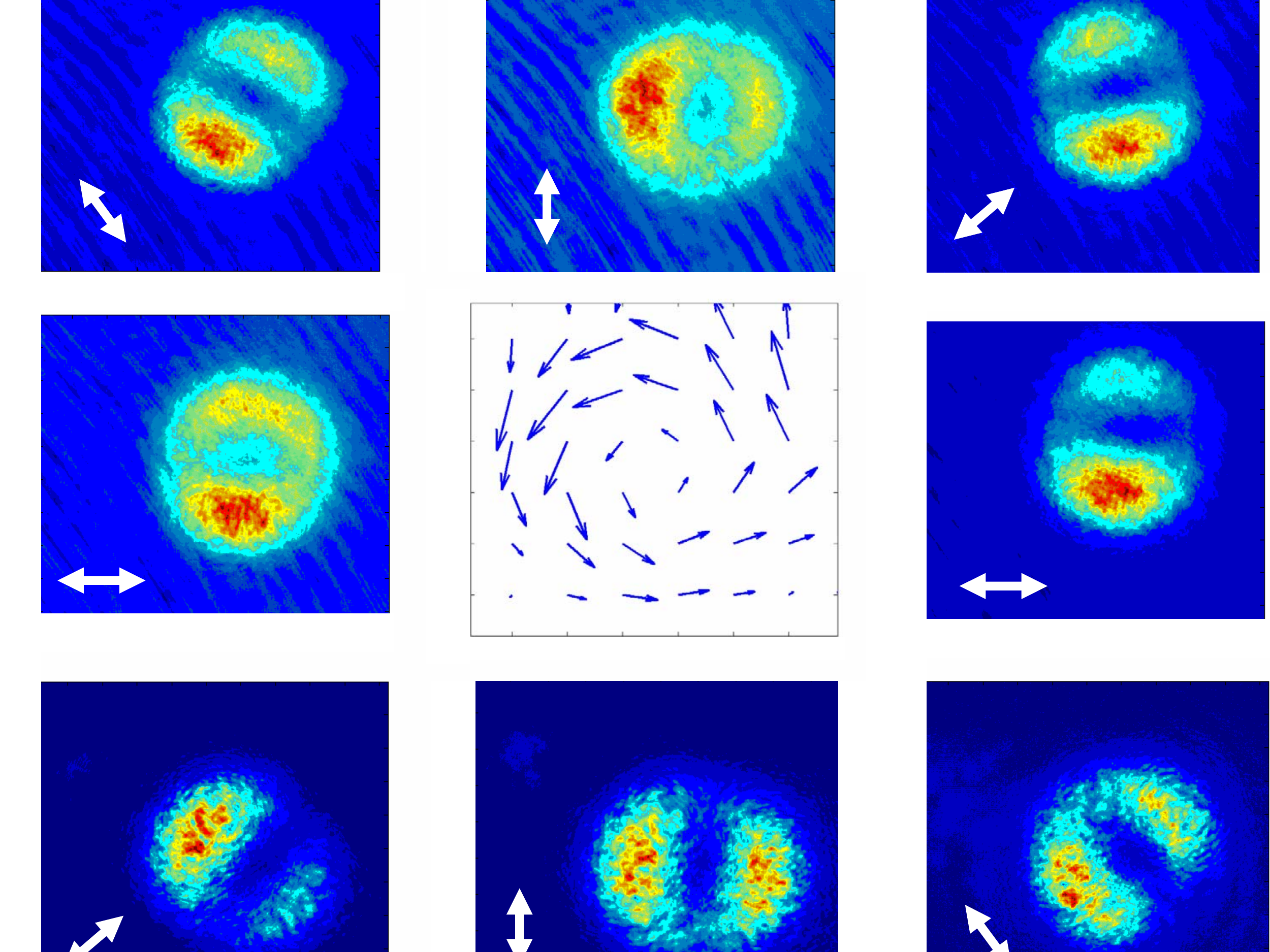}}
\caption{Representative cross-sections of the intensity
distributions of the output beam after the external linear
polarizer in eight different orientations when the SVR is oriented
at $45^{\circ}$. At the center, the resulting main axis of the
local polarization ellipse.} \label {azimuthal}
\end{figure}

Figure 4 shows the results when the SVR is oriented at
$45^{\circ}$. As evident from the detected distributions, the
orientation of the common bisector of the two lobes is normal to
the direction of the linear polarizer (denoted by the arrows),
indicating azimuthal polarization. Using these detected
distributions, we calculate the local polarization at any point
across the output beam. The results are shown in the center of
Fig. 4, where the arrows indicate the direction of the main axis
of the local polarization ellipse. We determined that the
azimuthal polarization purity was 90\%.

In order to compare our results of radially and azimuthally
polarized output light with those of linearly polarized light, we
removed the SVR shown in the configuration of Fig. 1 and inserted
an aperture to suppress all modes higher than the $TEM_{00}$ mode.
This resulted in our laser operating with a linearly polarized
Gaussian $TEM_{00}$ mode, where the polarization purity of the
output light was measured to be only 70\%. On the other hand, when
the laser operated with either radially or azimuthally polarized
light, the polarization purity was 90\% or better. These results
indicate that redial and azimuthal polarizations are more suitable
than linear polarization for multi-mode fiber lasers.

To conclude, we presented a simple and efficient method to produce
radially and azimuthally polarized light with 90\% or better
purity directly from fiber lasers. Although our experiments were
performed with relatively low output powers, we expect that our
method could be extended to higher powers which are needed in many
applications.

This research was supported in part by the Binational Science
Foundation.


\end{document}